\documentclass[12pt]{article}
\usepackage{cite}
\usepackage{textcomp}
\usepackage{amsmath,amssymb,amsfonts}
\usepackage{algorithmic}
\usepackage{graphicx}
\usepackage{textcomp}
\usepackage{xcolor}
\begin{document}
\sloppy 

{\huge\bf{Sonic shrinking of Pickering-stabilised ultrasound contrast agent at a low acoustic amplitude}}\\

Nicole Anderton$^1$\footnote[2]{Correspondence to: nicole.anderton@tuni.fi}, 
Craig Carlson$^2$,
Ryunosuke Matsumoto$^3$, 
Ri-ichiro Shimizu$^3$,  
Albert T. Poortinga$^4$, 
Nobuki Kudo$^3$, 
and Michiel Postema$^{1,2}$ \\
($^1$BioMediTech, Fac. Med. \& Health Technol., Tampere Univ., Finland; 
$^2$School Elec. \& Inform. Eng., Univ. Witwatersrand, Johannesburg, South Africa; 
$^3$Fac. Inform. Sci. \& Technol., Hokkaido Univ., Sapporo, Japan; 
$^4$Dept. Mech. Eng., Eindhoven Univ. Technol., Eindhoven, Netherlands)\\

{\it Submitted to the 42nd Symposium on UltraSonic Electronics (USE2021).}

\section{Introduction}
Ultrasound contrast agents comprise gas microbubbles surrounded by stabilising elastic or viscoelastic shells.$^{1-4}$ Microbubbles containing liquid or solid cores are referred to as antibubbles.$^5$ The manufacturing process of long-lived antibubbles involves the adsorption of colloidal particles at the interfaces, a process called Pickering stabilisation.$^6$

With and without cores present inside, Pickering-stabilised microbubbles generate a harmonic response, even at modest transmission amplitudes.$^7$ Therefore, Pickering-stabilised ultrasound contrast agents may be of interest in contrast-enhanced ultrasonic imaging.

In a previous study, we determined that the presence of a core inside Pickering-stabilised microbubbles slightly hampered the oscillation amplitude compared to identical microbubbles without a core.$^8$ The purpose of the present study is to determine whether the absence of a core negatively influences the stability of Pickering-stabilised microbubbles under sonication.

\section{Materials and Methods}
Pickering-stabilised microbubbles without cores (MB) and antibubbles with cores (AB) were produced as described by Poortinga,$^9$ with ZnO instead of aqueous cores and Aerosil\textsuperscript{\textregistered} R972 hydrophobised silica particles (Evonik Industries AG, Essen, Germany) as a stabilising agent. The preparation for the experiments was identical to the procedure of Kudo et al.$^5$

A 0.2-ml volume of either MB or AB suspension was pipetted into the observation of a high-speed observation system.$^10$ The observation chamber was placed under an IX70 microscope (Olympus Corporation, Shinjuku-ku, Tokyo, Japan) with a LUMPlan FI/IR 40$\times$ (N.A. 0.8) objective lens. The microscope was coupled to an HPV-X2
high-speed camera (Shimadzu, Nakagyo-ku, Kyoto, Japan), operating at a recording speed of 10 million frames per second with exposure times of 100 ns per frame. High-speed videos were recorded during sonication. The sonication equipment was described by Kudo et al.$^8$
A burst comprised a 3-cycle sine pulse with a centre frequency of 1 MHz and a peak-negative pressure of 0.2 MPa, which corresponds to a low mechanical index of 0.2. The pulse started in compression phase.

Each video recorded consisted of 256 still frames. The frames were clipped, segmented, and analysed using MATLAB\textsuperscript{\textregistered} (The MathWorks, Inc., Natick, MA, USA). For individual MB and AB, the radius was determined as a function of time.

A total number of 112 MB and 158 AB were included in this study. The size distributions of both populations before sonication are shown in \textbf{Fig. 1}, where the blue and black curves represent the resting radii $R_0$ of MB and AB, respectively. The median resting radius of MB was $R_0=2.3~\mu$m and of AB $R_0=3.0~\mu$m.
\begin{figure}[htbp]
\centerline{
\includegraphics[width=1.0\linewidth]{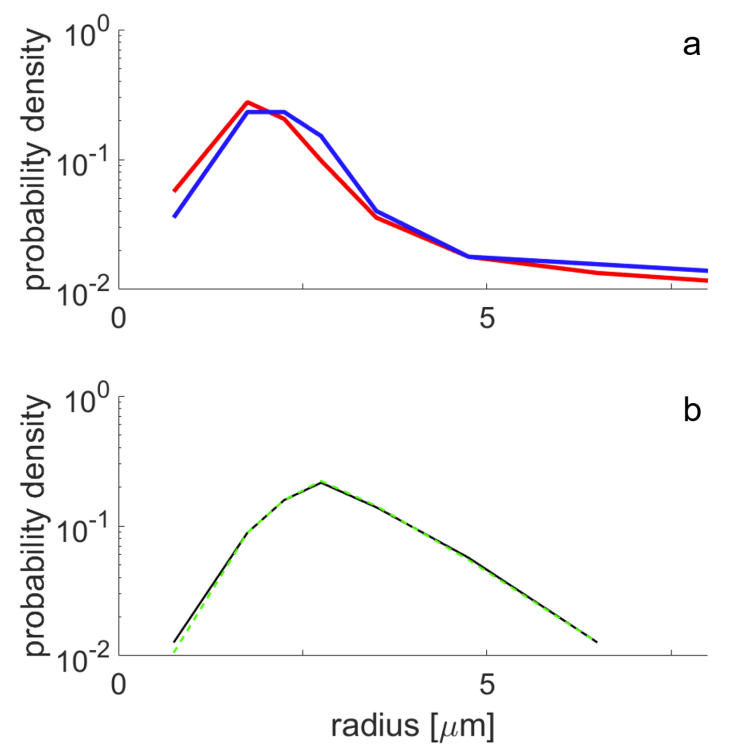}
}
\caption{Size distribution of MB (a), before (blue) and after (red) sonication; size distribution of AB (b), before (black) and after (green) sonication.} 
\end{figure}

\begin{figure}[htbp]
\includegraphics[width=1.0\linewidth]{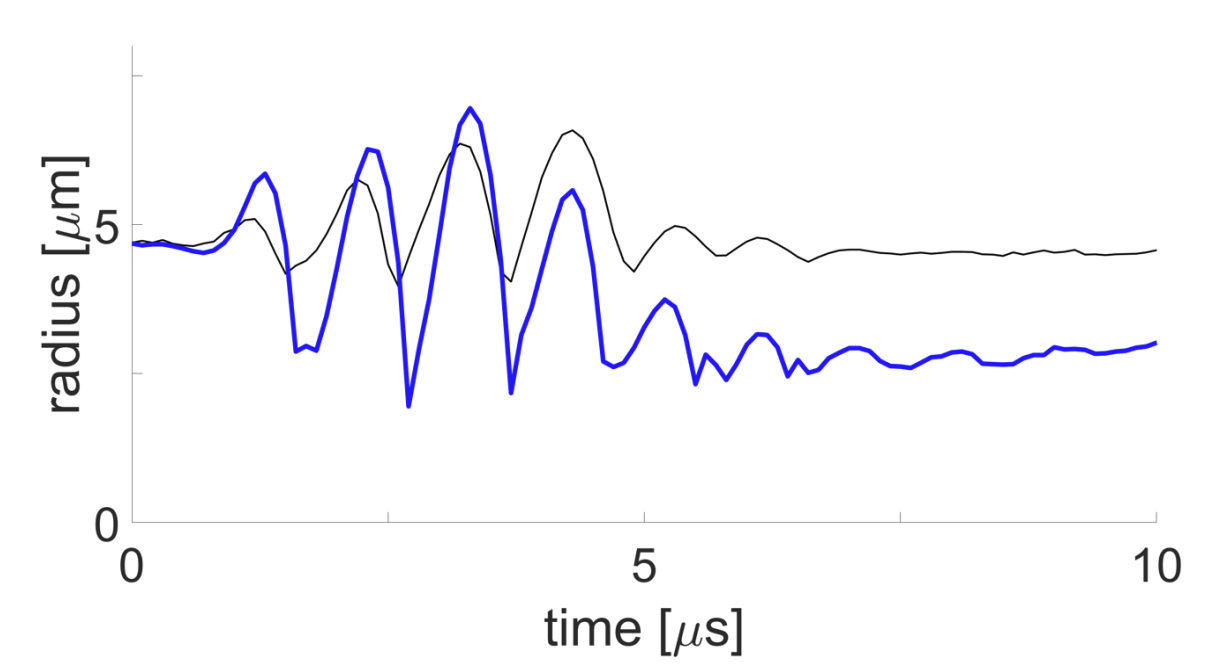}
\caption{Radius as a function of time of MB (blue) and AB (black).}
\end{figure}

\section{Results and Discussion}
Examples of two radius(time) curves are shown in \textbf{Fig. 2}. The resting radii of both MB and AB before sonication were $R_0=4.7~\mu$m.

MB was observed to shrink during sonication. The final radius after sonication was measured to be $R\textsubscript{fin} = 2.9~\mu$m. Although the oscillation amplitude of AB approximated that of MB, it was not observed to shrink. Its final radius after sonication was measured to be $R\textsubscript{fin} = 4.6 \mu$m, a negligible difference with $R_0$.

The size distributions of all MB and AB measured after sonication have been added to Fig. 1. The median final radius of MB had decreased to $R\textsubscript{fin} = 2.1~\mu$m, whilst the median radius of AB had remained $R\textsubscript{fin} = 3.0~\mu$m$\, = R_0$.

A full overview of final radius as a function of resting radius is shown in \textbf{Fig.\,3}. Linear regression for MB yielded $R\textsubscript{fin} = 0.98 R_0 – 0.15 \mu$m, indicated by the blue line. Linear regression for AB yielded $R\textsubscript{fin}=1.0 R_0 – 0.012 \mu$m (not shown), i.e., shrinkage very much below significance. The reasons why the absence of cores leads to shrinking cannot be determined from the video footage. Nevertheless, the speed of the process suggests that physical gas release from MB, also referred to as sonic cracking, is involved. Sudden gas release might generate a short, transient acoustic response.

We observed a 0.2-$\mu$m median shrinking in the bulk of MB. As this must be accompanied by a slight increase in the resonance frequency of the population, it is interesting to think of diagnostic applications of the change in acoustic signature.
\begin{figure}[htbp]
\centerline{
\includegraphics[width=1.0\linewidth]{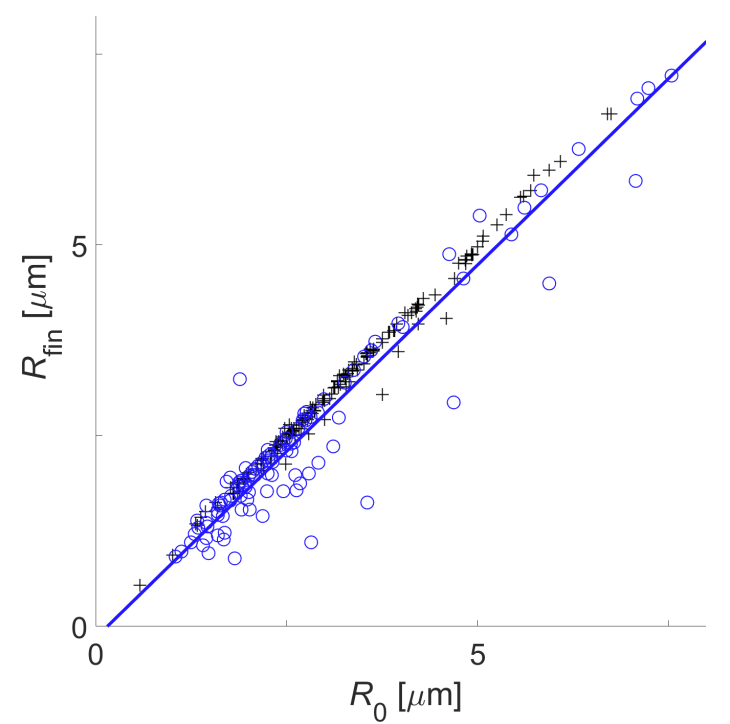}
}
\caption{Final radius $R\textsubscript{fin}$ as a function of initial resting radius $R_0$ for MB (o) and AB (+). Linear regression for MB yielded $R\textsubscript{fin} = 0.98 R_0 –0.15 ~\mu$m (blue line).} 
\end{figure}

\section{Conclusions}
It is concluded that Pickering-stabilised microbubbles without cores are less stable under the acoustic conditions used than those containing cores.

Sonic shrinking must increase the resonance frequency of the microbubbles. Therefore, the proven instability of the Pickering-stabilised
microbubbles without cores may create interesting transient responses in ultrasound at diagnostic amplitudes.

\section*{Acknowledgments}
This work was supported by JSPS KAKENHI, Grant Nos. JP17H00864 and JP20H04542, by the National Research Foundation of South Africa, Grant No. 127102, and by the Academy of Finland, Grant No. 340026.

\section*{References}
\begin{enumerate}
\item	M. Chan and K. Soetanto: Jpn. J. Appl. Phys. \textbf{37} (1998) 3078.
\item	H. Yoshikawa et al.:Jpn. J. Appl. Phys. \textbf{45} (2006) 4754.
\item	S. Park et al.: Jpn. J. Appl. Phys. \textbf{56} (2017) 07JF10.
\item	Y. Kikuchi and T. Kanagawa: Jpn. J. Appl. Phys. \textbf{60} (2021) SDDD14.
\item	N. Kudo et al.: Jpn. J. Appl. Phys. \textbf{59} (2020) SKKE02.
\item	A. T. Poortinga: Langmuir \textbf{27} (2011) 2138.
\item   M. Postema et al.: Appl. Acoust. \textbf{137} (2018) 148.
\item   N. Kudo: Proc. Symp. UltraSon. Electron. \textbf{40} (2019) 2E3-1.
\item   A. T. Poortinga: Colloids Surf. A \textbf{419} (1013) 15.
\item   N. Kudo: IEEE Trans. Ultrason. Ferroelect. Freq. Control \textbf{64} (2017) 273.
\end{enumerate}

\end{document}